\documentclass[sigconf]{acmart} %% The correct template for publication

\AtBeginDocument{%
  }

\copyrightyear{2023}
\acmYear{2023}
\setcopyright{rightsretained}
\acmConference[CHI '23]{Proceedings of the 2023 CHI Conference on Human Factors in Computing Systems}{April 23--28, 2023}{Hamburg, Germany}
\acmBooktitle{Proceedings of the 2023 CHI Conference on Human Factors in Computing Systems (CHI '23), April 23--28, 2023, Hamburg, Germany}\acmDOI{10.1145/3544548.3580882}
\acmISBN{978-1-4503-9421-5/23/04}

\begin{document}

\title{Understanding Frontline Workers’ and Unhoused Individuals’ Perspectives on AI Used in Homeless Services}


\author{Tzu-Sheng Kuo}
\authornote{Co-first authors contributed equally to this research.}
\email{tzushenk@cs.cmu.edu}
\affiliation{
  \institution{Carnegie Mellon University}
  \city{Pittsburgh}
  \state{PA}
  \country{USA}
}

\author{Hong Shen}
\authornotemark[1]
\email{hongs@cs.cmu.edu}
\affiliation{
  \institution{Carnegie Mellon University}
  \city{Pittsburgh}
  \state{PA}
  \country{USA}
}

\author{Jisoo Geum}
\email{geumjisoo@gmail.com}
\affiliation{
  \institution{Carnegie Mellon University}
  \city{Pittsburgh}
  \state{PA}
  \country{USA}
}

\author{Nev Jones}
\email{nevjones@pitt.edu}
\affiliation{
  \institution{University of Pittsburgh}
  \city{Pittsburgh}
  \state{PA}
  \country{USA}
}

\author{Jason I. Hong}
\email{jasonh@cs.cmu.edu}
\affiliation{
  \institution{Carnegie Mellon University}
  \city{Pittsburgh}
  \state{PA}
  \country{USA}
}

\author{Haiyi Zhu}
\authornote{Co-senior authors contributed equally to this research.}
\email{haiyiz@cs.cmu.edu}
\affiliation{
  \institution{Carnegie Mellon University}
  \city{Pittsburgh}
  \state{PA}
  \country{USA}
}

\author{Kenneth Holstein}
\authornotemark[2]
\email{kjholste@cs.cmu.edu}
\affiliation{
  \institution{Carnegie Mellon University}
  \city{Pittsburgh}
  \state{PA}
  \country{USA}
}

\renewcommand{\shortauthors}{Kuo and Shen et al.}

\begin{abstract}

Recent years have seen growing adoption of AI-based decision-support systems (ADS) in homeless services, yet we know little about stakeholder desires and concerns surrounding their use. In this work, we aim to understand impacted stakeholders’ perspectives on a deployed ADS that prioritizes scarce housing resources. We employed AI lifecycle comicboarding, an adapted version of the comicboarding method, to elicit stakeholder feedback and design ideas across various components of an AI system’s design. We elicited feedback from county workers who operate the ADS daily, service providers whose work is directly impacted by the ADS, and unhoused individuals in the region. Our participants shared concerns and design suggestions around the AI system’s overall objective, specific model design choices, dataset selection, and use in deployment. Our findings demonstrate that stakeholders, even without AI knowledge, can provide specific and critical feedback on an AI system’s design and deployment, if empowered to do so.

\end{abstract}





\begin{teaserfigure}
  \includegraphics[width=\textwidth]{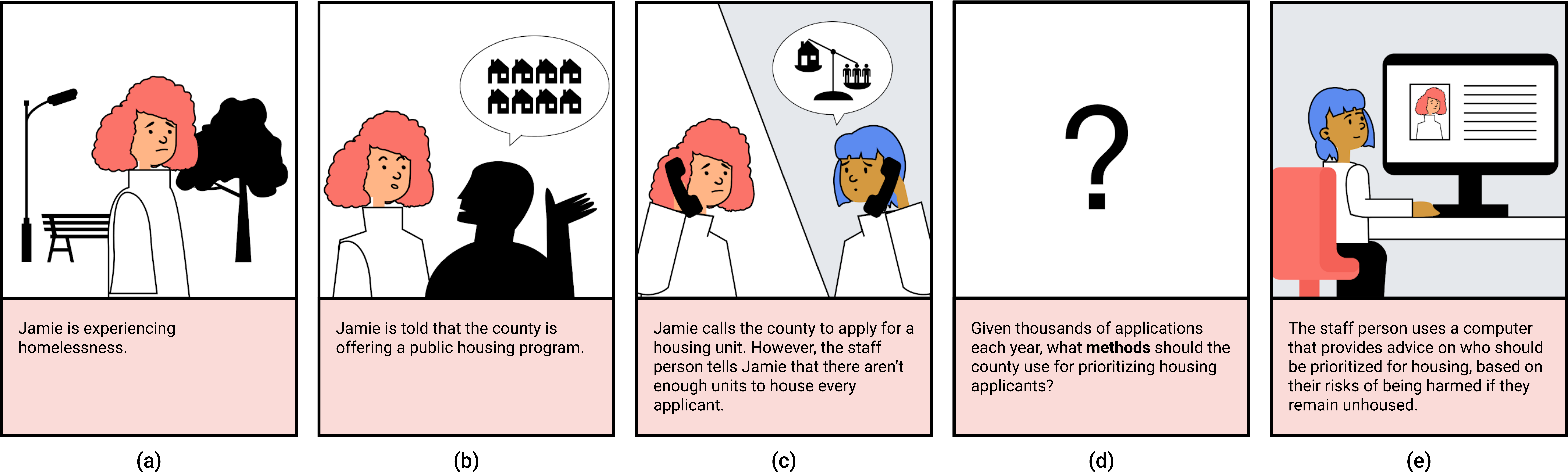}
  \caption{We use an adapted version of the comicboarding method \cite{moraveji2007comicboarding} to understand frontline workers' and unhoused individuals' perspectives on an AI system used in homeless service. In order to elicit specific stakeholder feedback and design ideas around the design and deployment of an AI system, our adaptation disaggregates the AI development lifecycle into different design components of an AI system, such as the system's task definition (as shown in this example).}
  \Description{This figure is one of the comicboards we developed. This comicboard has five panels, each of which includes an illustration and a brief caption. The first panel says: Jamie is experiencing homelessness. The second panel says: Jamie is told that the county is offering a public housing program. The third panel says: Jamie calls the county to apply for a housing unit. However, the staff person tells Jamie that there aren’t enough units to house every applicant. The fourth panel asks: Given thousands of applications each year, what methods should the county use for prioritizing housing applicants? The last panel says: The staff person uses a computer that provides advice on who should be prioritized for housing, based on their risks of being harmed if they remain unhoused.}
  \label{fig:teaser}
\end{teaserfigure}

\maketitle

\section{Introduction}

According to the United Nations, homelessness is a ``profound assault on dignity, social inclusion and the right to life’’ \cite{un2019guidelines}. More than 1.8 billion people lack adequate housing worldwide \cite{un2019guidelines}. Even in developed countries such as the United States, more than 326,000 people experienced sheltered homelessness on a single night in 2021 \cite{Henry2022}. This number does not even account for those experiencing unsheltered homelessness and growth in the unhoused population due to the global pandemic \cite{Henry2022, pixley2022role}. Furthermore, homelessness is often deeply stigmatized, with specific stereotypes linking unhoused individuals with dangerousness, criminality, and moral failure \cite{phelan1997stigma, kim2021analyzing}. Public prejudice further marginalizes some of the most vulnerable in our society \cite{kurzban2001evolutionary}.

In recent years, government agencies have increasingly turned to AI-based decision-support (ADS) systems to assist in prioritizing scarce housing resources. The use of algorithmic systems in homeless services has spread rapidly: in the past half-decade, such systems have been considered or deployed in US counties including Los Angeles \cite{Denton2019}, San Francisco \cite{Thompson2021}, Allegheny County \cite{pittsburgh2020report}, as well as in Ontario, Canada \cite{Lamberink2020}. However, despite this rapid spread, the stakeholders most directly impacted by these systems have had little say in these systems’ designs \cite{Denton2019, Thompson2021}. To date, we lack an adequate understanding of impacted stakeholders’ desires and concerns around the design and use of ADS in homeless services. Yet as a long line of research in HCI and participatory design demonstrates, without such an understanding to guide design, technology developers risk further harming already vulnerable social groups, or missing out on opportunities to better support these groups \cite{costanza2020design, eubanks2018automating, delgado2021stakeholder, stapleton2022imagining, woodruff2018qualitative}.

In this work, we aim to understand frontline workers’ and unhoused individuals’ perspectives on the \textit{Housing Allocation Algorithm (HAA)}\footnote{Throughout the paper, we refer to the ADS system by this pseudonym.}, an ADS system that prioritizes housing resources for people experiencing homelessness, which has been deployed in a US county for over two years. We employ \textit{AI lifecycle comicboarding}, a feedback elicitation and co-design method adapted from comicboarding \cite{hiniker2017co, moraveji2007comicboarding}, to elicit specific feedback on various aspects of an AI system's design from stakeholders with diverse backgrounds and literacies. Prior HCI methods aimed at broadening participation in the design and critique of AI systems have often focused either on eliciting broad feedback on AI systems’ overall designs (e.g., feedback on the problem framing and design objectives) \cite{holten2020shifting, stapleton2022imagining, woodruff2018qualitative, veale2018fairness}, or on eliciting specific feedback by narrowing down the elicitation process around specific aspects of the system design \cite{lee2019webuildai, chen2022perspectives, cheng2021soliciting, robertson2020if, shen2022model}. Our adaptation uses comicboards (see Figure \ref{fig:teaser}) to scaffold both broad and specific conversations around different components of an AI system’s design, from problem formulation to data selection to model definition and deployment.

Using our adapted approach, we elicited feedback on HAA’s design from frontline workers, including county workers who operate the ADS daily and external service providers whose work is directly impacted by the ADS, as well as both current and former unhoused individuals in the region. We recruited these stakeholder groups to center the voices of those who are most directly affected by the system, yet who are currently least empowered to shape its design. Our participants shared critical concerns and specific design suggestions related to the system’s overall \textit{design objective}, specific \textit{model design} choices, the selection of \textit{data} use to train and operate the system, and broader \textit{sociotechnical design considerations} around the system’s deployment. Reflecting on their experience during the study, participants noted that our approach helped to open up conversations around otherwise hidden assumptions and decisions underlying the ADS’s design and deployment.

In summary, our work contributes an in-depth understanding of frontline workers’ and unhoused individuals’ perspectives on the use of AI in homeless services. In order to understand their perspectives, we employ AI lifecycle comicboarding, an adapted comicboarding method uniquely tailored to understand the sociotechnical implications of AI systems. Our adaptation disaggregates the AI development lifecycle to make otherwise opaque AI design choices accessible and open to critique. Overall, our findings demonstrate that stakeholders spanning a broad range of relevant literacies can provide specific and critical feedback on an AI system’s design and deployment, if empowered to do so.

\section{Related Work}

\subsection{ADS Systems in the Public Sector} \label{relatedwork-1}
Over the past decade, AI-based decision support (ADS) systems, powered by machine learning (ML) techniques, have increasingly been adopted to augment decision-making across a range of public services \cite{levy2021algorithms}. For example, ADS systems have been used to assist judges in deciding whether defendants should be detained or released while awaiting trial \cite{corbett2017algorithmic, dressel2018accuracy}. They have been adopted by child protection agencies to assist workers in screening child maltreatment referrals \cite{chouldechova2018case, kawakami2022improving, saxena2020human}. They have also been used by school districts to assist in assigning students to public schools \cite{robertson2021modeling}. The growing use of ADS in public services has been met with both enthusiasm and concern \cite{chouldechova2018case, levy2021algorithms}. While proponents have argued for its potential to improve the equity, efficiency and effectiveness of decision-making, critics have raised serious concerns about ways these systems may fail to deliver on these promises in practice, and instead amplify the problems that they were meant to address. For example, public outcry has erupted over biased and harmful outcomes caused by recidivism prediction algorithms \cite{angwin2016machine}, predictive analytics for child welfare \cite{eubanks2018automating}, and predictive policing \cite{shapiro2017reform}. 

A growing body of work in ML and HCI has begun to both deepen our understanding of stakeholder concerns and implement changes in AI systems in an effort to address them. Past work in the fair ML research community has focused on proposing various statistical fairness criteria and then developing novel algorithmic methods to align ML models with these criteria \cite{corbett2018measure, chouldechova2018frontiers, mehrabi2021survey}. However, early work in this space has been critiqued for its disconnectedness from real-world stakeholders' actual needs, values, and system constraints, which may not align with theoretical notions of fairness \cite{cheng2021soliciting, holstein2019improving, veale2018fairness}. Meanwhile, HCI research has paid increasing attention to understanding stakeholders’ desires and concerns around the design of public sector algorithms \cite{holten2020shifting, kawakami2022improving, robertson2021modeling, saxena2021framework, stapleton2022imagining, veale2018fairness}. For example, in the context of public education, Robertson et al. \cite{robertson2021modeling} found that student assignment algorithms deployed in San Francisco failed because they relied on modeling assumptions that fundamentally clashed with families' actual priorities, constraints, and goals.  

In contrast to the domains discussed above, the rapid spread of ADS in homeless services has received surprisingly limited attention from the HCI community to date. Yet the lack of engagement of directly impacted stakeholders in this high-stakes domain has already received attention in the popular press, regarding the potential for such systems to negatively impact already vulnerable social groups (e.g., \cite{Denton2019, Thompson2021}).

\subsection{Broadening Participation in AI Design} \label{relatedwork-2}
In current practice, stakeholders with no background in statistics or data science and even researchers and designers who have less technical knowledge of AI and machine learning are often excluded from conversations around the design and use of AI-based systems \cite{delgado2021stakeholder, holten2020shifting, kawakami2022improving}. Directly impacted stakeholders–e.g. individuals facing criminal adjudication, children and families subject to out of home placements and state intervention,  individuals with intellectual/developmental disabilities–are even less likely to be meaningfully included in algorithm development than other groups. Although top down development can reduce the burden on AI teams, in terms of having to explain complex technical concepts to laypeople, a lack of feedback from stakeholders who will use or be directly impacted by these systems can lead to serious misalignments and failures once these systems are deployed (e.g., \cite{kawakami2022improving, robertson2021modeling, saxena2021framework, stapleton2022imagining, veale2018fairness}). Even when less technical stakeholders are involved in design conversations, they may not be provided with sufficiently detailed information about the system to offer specific critiques and suggestions, or they may only be invited to provide feedback on a narrow aspect of an AI system’s design. For example, stakeholders might be invited to critique the design of a system’s user interface, but not the underlying model or the overall problem formulation \cite{delgado2021stakeholder, yang2018investigating}.

In response, a growing body of research in HCI and ML has focused on the development of new methods and tools aimed at broadening who is able to participate in the design or critique of AI-based systems \cite{delgado2021stakeholder, halfaker2020ores, holstein2020replay, kulynych2020participatory, shen2022model, zhu2018value, zytko2022participatory}. A body of prior work has focused on inviting high-level feedback on AI systems’ overall designs (e.g., feedback on the problem framing and design objectives), through a wide range of methods including interviews, workshops, and co-design techniques (e.g.,\cite{holten2020shifting, stapleton2022imagining, woodruff2018qualitative, veale2018fairness}). The feedback that results from these approaches can be extremely valuable in informing research and design teams’ understandings of stakeholders’ broad desires and concerns around specific kinds of AI systems. For example, recent work from Stapleton et al. found that both community members and social workers \cite{stapleton2022imagining} had major concerns about existing uses of predictive analytics in child welfare decision-making, and desired fundamentally different forms of technology-based support. However, these approaches often fail to make explicit the assumptions and design choices involved in an AI system's design (e.g., specific choices of training data or proxy outcomes that a model predicts), and thus fail to provide the knowledge and context that would be necessary to elicit more specific critiques.

By contrast, another body of work has focused on developing methods and tools that can elicit more detailed stakeholder feedback, by narrowing down the elicitation process around specific aspects of the system design (e.g., \cite{lee2019webuildai, chen2022perspectives, cheng2021soliciting, shen2022model}). For example, Lee et al. \cite{lee2019webuildai} developed a voting-based preference elicitation approach to solicit stakeholders preferences regarding how a matching algorithm should prioritize stakeholder groups. Such approaches are powerful in eliciting stakeholder feedback that can directly inform the design or redesign of AI systems. However, as discussed in recent work \cite{robertson2020if}, these approaches also risk limiting the types of feedback that stakeholders are able to provide, by restricting their inputs to forms that are readily computable. For example, a method might ask stakeholders to weigh in on how an algorithm should prioritize services among different groups, without providing opportunities for them to reflect on whether the proposed algorithm is in fact addressing the right problem in the first place \cite{robertson2020if}. 

In this paper, we introduce an adaptation of the comicboarding method \cite{hiniker2017co, moraveji2007comicboarding}, which aims to complement these prior approaches by scaffolding participants to provide targeted feedback on key components of an AI system's design, from a system's problem formulation to the selection of training data to the design of a model and its use in deployment.

\section{Study Context} \label{studycontext}

\begin{figure*}[!t]
  \centering
  \includegraphics[width=0.94\linewidth]{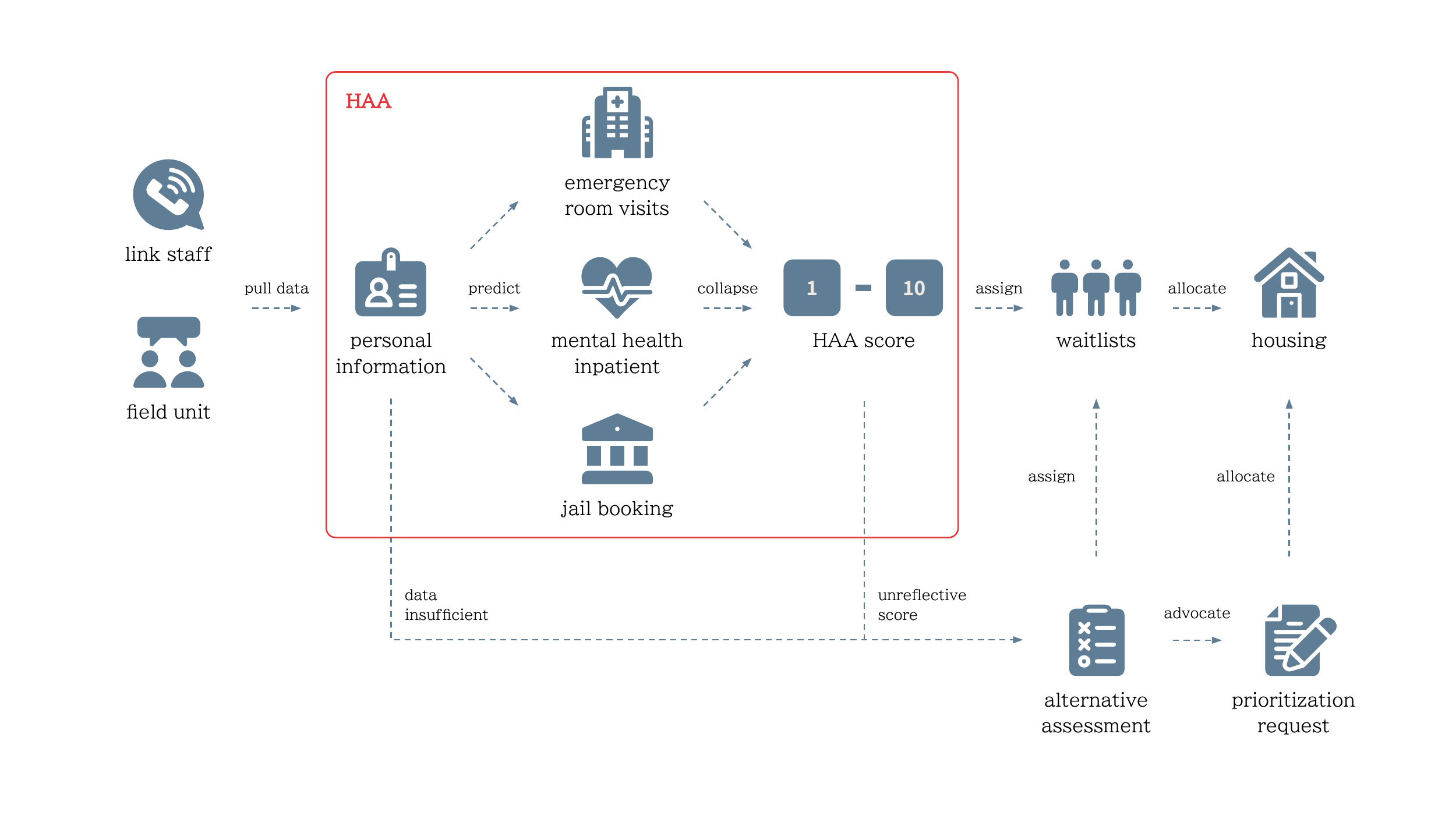}
  \caption{Diagram showing HAA’s workflow. HAA automatically pulls an applicant's personal information from the county's data warehouse, predicts an applicant’s likelihood of adverse events such as emergency room visits, and generates a score that determines an applicant's position on the housing waitlists.}
  \Description{This figure is a diagram that describes HAA's workflow. It has a graphical icon for each of the following components mentioned in the section on study context: link staff, field unit, personal information, emergency room visits, mental health inpatient, jail booking, HAA score, waitlists, housing, alternative assessment, and prioritization requests. Several arrows are pointing between these icons, mostly from left to right. An arrow points from link staff and field unit toward personal information. Three arrows point from personal information toward each of the following, emergency room visits, mental health inpatient, and jail booking. Another three arrows point from each of these three events toward HAA score. An arrow points from HAA score to waitlists. An arrow points from waitlists to housing. Two arrows point from personal information and HAA score to alternative assessment, which has another two arrows from which point toward waitlists and prioritization request. Finally, an arrow points from prioritization request toward housing.}
  \label{workflow}
\end{figure*}

We conducted this research in the United States, where the unhoused population has grown significantly in the past decades \cite{shlay1992social}, and the use of algorithmic systems in homeless services has rapidly expanded \cite{Denton2019, Thompson2021, pittsburgh2020report}. Following the definition used by the U.S. Department of Housing and Urban Development, throughout this paper, we consider the experience of homelessness as the lack of ``a fixed, regular, and adequate nighttime residence'' \cite{Henry2022}.

Prior work in HCI has studied the perceptions, uses, and impacts of technologies among both people experiencing homelessness and homeless service providers in the United States. For example, early work by Le Dantec et al. \cite{le2008designs, le2011publics} and Roberson et al. \cite{roberson2010survival} explored how commodity technologies, such as mobile phones, affect unhoused individuals' daily lives. With the rise of social media, HCI researchers investigated how unhoused individuals used social media to portray life on the streets, develop social ties, and meet survival needs \cite{woelfer2010homeless, hu2019characterizing, woelfer2012homeless}. More recently, given the rapid spread of data-driven tools used for housing allocation, Karusala et al. conducted semi-structured interviews with policymakers and homeless service providers to understand data practices around a questionnaire-based triage tool, and to understand participants' desires for new potential uses of these data \cite{karusala2019street}. 

Our work builds upon this line of HCI research, centering the voices of unhoused individuals and frontline workers in homeless services. In this research, we focus on a US county where an ADS for housing resource allocation has been in use for over two years. According to the county, the housing units available through turnover can serve fewer than half of the individuals or families experiencing homelessness in the county, resulting in a housing gap of more than a thousand people a year \cite{AHA2020faq}. Due to the limited housing availability, the county began exploring ways to improve its assessment tool for housing prioritization about five years prior to our study \cite{AHA2020faq}. At that time, the county used a questionnaire-based assessment tool \cite{AHA2021bumpyride}. However, according to the county \cite{AHA2020faq, AHA2020focusgroup}, these questionnaires were time consuming for applicants to fill out, and the information gathered was potentially inaccurate, given that the questionnaire required applicants to answer highly sensitive questions. For example, a question asked applicants whether they have ``exchange[d] sex for money, run drugs for someone, or have unprotected sex with someone you don’t know’’ \cite{AHA2020faq}. Furthermore, the county worried that answering such sensitive questions risked retraumatizing applicants, who might be forced to relive difficult experiences while completing the questionnaire \cite{AHA2020faq}. To address these concerns, the county collaborated with external researchers to develop an ADS that prioritizes housing resources for people experiencing homelessness using government administrative records instead of questionnaire responses \cite{AHA2020methodology, AHA2020update, AHA2020eticas}. In the remainder of this paper, we refer to this ADS system with the pseudonym ``HAA,’’ an abbreviation of the \textit{Housing Allocation Algorithm}.

HAA’s workflow is illustrated in Figure \ref{workflow}. The housing assessment process starts with an applicant getting in contact with either a county’s link staff in the office or a frontline worker in the field unit. After the county staff or worker asks a few filtering questions to ensure that the applicant is eligible for the assessment, they press a button on their computers to run HAA. HAA automatically pulls the applicant's personal information from the county's data warehouse \cite{AHA2021datawarehouse} and predicts how likely the applicant will experience the following three situations if they remain unhoused over the next 12 months: more than four emergency room visits based on healthcare utilization data, at least one mental health inpatient funded by Medicaid, and at least one jail booking. Based on this likelihood, HAA generates a risk score between 1 to 10. The higher the score, indicates the higher risk of being harmed due to homelessness. The staff or worker then assigns applicants to the housing waitlists corresponding to their scores. Once these housing programs have openings, the county connects people with the downstream housing providers.

Sometimes, county workers run the alternative assessment \cite{AHA2020methodology}, which is an adaptation of the previous questionnaire-based assessment tool that relies on an applicant’s self-reported information. In the remainder of this paper, we refer to this alternative assessment with the pseudonym ``alt HAA,’’ an abbreviation of the \textit{alternative Housing Allocation Algorithm}. The county uses alt HAA under two circumstances \cite{AHA2020faq}. First, when an applicant's data within the county exists for less than 90 days and thus has no sufficient data for the system to pull. Second, when workers believe that HAA’s score doesn't reflect an applicant's vulnerability. The alt HAA generates an alternative score, also between 1 to 10, that the county uses to assign applicants to waitlists. In rare cases, county workers may file a prioritization request when they believe neither HAA's nor alt HAA's score reflects an applicant's vulnerability.

\section{Study Design}

Drawing upon prior HCI methods, we introduced ``AI lifecycle comicboarding,’’ an adaptation of the comicboarding method discussed below. Using this method, we conducted a series of one-on-one study sessions with unhoused individuals and frontline workers, to understand their perspectives, concerns, and desires around the use of AI in homeless services. As discussed below, we also presented participants with de-identified design ideas and feedback generated by prior participants from the two other stakeholder groups, in order to facilitate asynchronous inter-group deliberation without compromising participant comfort and safety during our study. 

\subsection{AI Lifecycle Comicboarding}

\begin{figure*}[!ht]
  \centering
  \includegraphics[width=\linewidth]{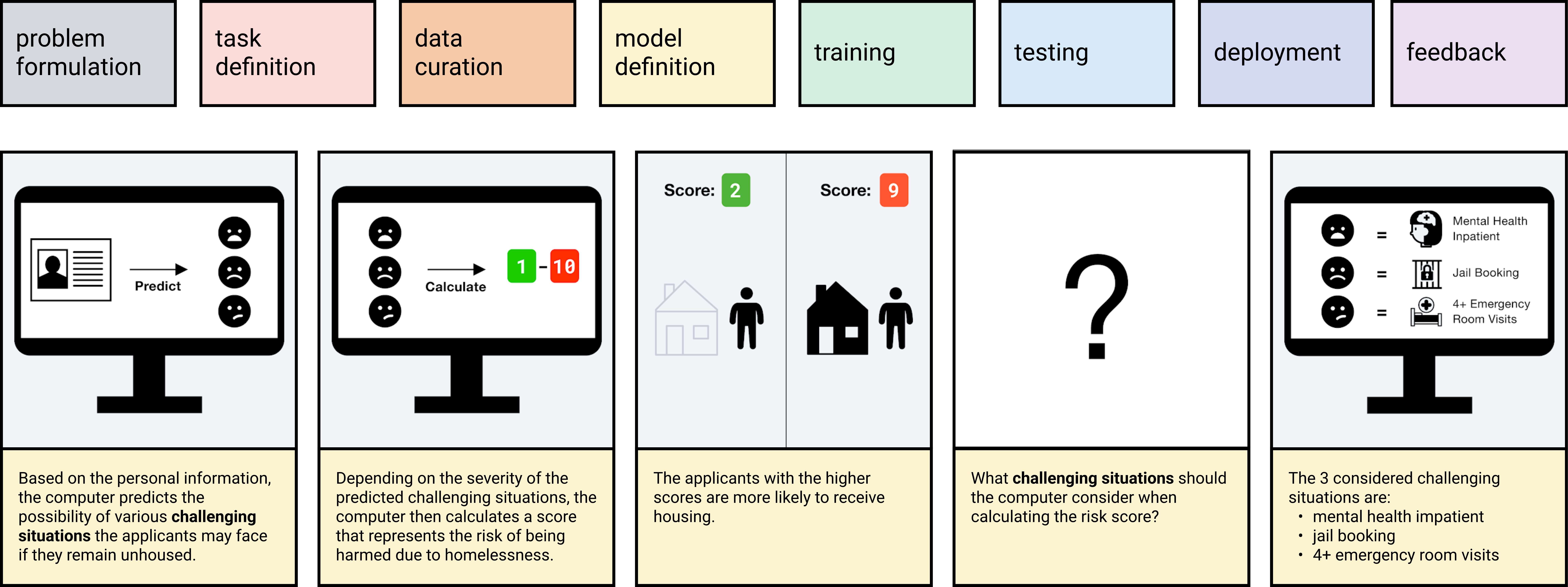}
  \caption{We developed a set of comicboards based on eight major components of an AI system's design \cite{cramer2019challenges}, as shown in the top row. We then used these comicboards as scaffolding to elicit participant's feedback around each component. For example, the pictured comicboard was created for the \textit{model definition} component, based on detailed technical reports released by the county about HAA \cite{AHA2020methodology, AHA2020update, AHA2020eticas}.}
  \Description{This figure has two rows. The upper row has eight major components of an AI system's design. These eight components, listed from left to right, are problem formulation, task definition, data curation, model definition, training, testing, deployment, and feedback. The bottom row is a comicboard we developed for the model definition component. There are five panels in this comicboard. The first panel says: Based on the personal information, the computer predicts the possibility of various challenging situations the applicants may face if they remain unhoused. The second panel says: Depending on the severity of the predicted challenging situations, the computer then calculates a score that represents the risk of being harmed due to homelessness. The third panel says: The applicants with the higher scores are more likely to receive housing. The fourth panel asks: What challenging situations should the computer consider when calculating the risk score? The final panel says: The three considered challenging situations are: mental health impatient, jail booking, and 4+ emergency room visits.}
  \label{lifecycle}
\end{figure*}

We adapted the comicboarding method \cite{hiniker2017co, moraveji2007comicboarding} with the goal of scaffolding conversations around various aspects of an AI system’s design and deployment, when working with participants who span a broad spectrum of relevant reading and technology literacies (in the context of our study: unhoused individuals and frontline workers in homeless services). Comicboarding is a co-design technique that uses the structure of comic strips, including partially completed content, as scaffolding to facilitate idea generation and elicit design insights from populations who may have little experience with brainstorming. The comicboarding method has been successfully used in prior research to elicit rich design ideas and feedback from participants with lower reading literacy \cite{hiniker2017co, moraveji2007comicboarding}. This property made comicboarding a strong fit for our context, given that low reading literacy presents a barrier to participation in co-design for many unhoused participants \cite{grajo2020effectiveness}. 

We adapted this method to engage diverse stakeholders in conversations around the design and use of AI systems. In order to surface AI design choices and assumptions that might otherwise remain opaque to non-AI experts, we developed targeted comicboards to solicit design feedback and ideas around various aspects of an AI system’s design. We created a comicboard for each of eight major components of an AI system’s design, each corresponding to a different stage of the AI development lifecycle \cite{cramer2019challenges}: problem formulation, task definition, data curation, model definition, training, testing, deployment, and feedback processes. Each of these eight comicboards consists of a set of ``story starter’’ panels, which set the context for a key design choice (e.g., see Figure \ref{lifecycle} for an example in the context of \textit{model definition}). The next panel in a comicboard then presents a blank space, along with an open-ended question to prompt ideation and critique. The final comicboard panel, which is revealed to a participant only after they have engaged around the open-ended panel, reveals how the given aspect of the system is \textit{currently} designed in reality (or the proposed design, in the case of systems that have not yet been developed).

Using this approach, in the current study we developed a set of comicboards to understand frontline workers' and unhoused populations' perspectives on AI used in homeless services. All our comicboards are shown in Appendix \ref{comicboards}. We based the information in each of our comicboards upon detailed technical reports published by the county about the HAA system \cite{AHA2020methodology, AHA2020update, AHA2020eticas}. As such, our comicboards served to surface key design decisions that were buried in these lengthy reports, translating this information into a format that could be more readily scrutinized by our participants. For example, our comicboard for \textit{task definition} invited design ideas regarding the AI system's overall objective; our comicboard for \textit{model definition} elicited feedback on the choice of proxies that the model used to predict a person's risk of being harmed due to homelessness. In addition to the key design decisions we asked participants about in the blank comicboard panels, we also embedded various design choices throughout the comicboards to elicit specific feedback on each component. For example, in the comicboard for \textit{data curation}, we surfaced details about the training data to elicit discussion around potential issues surrounding its representativeness of the local unhoused population. In the comicboard for \textit{feedback}, we illustrated the third-party consulting firm from which the county gathers technical feedback in order to elicit responses to the values encoded in the current feedback process.

Finally, through several meetings with the county, we validated that our comicboards accurately reflected HAA’s current design and use. Given the stigma surrounding issues of homelessness, we used a gender neutral persona throughout the storyboards, so that study participants had the option to refer to the persona in third-person when discussing sensitive topics, rather than directly referencing their own experiences.

\begin{table*}
  \caption{Potential barriers to participation in co-design and design critique in the context of ADS used in homeless services, and how our approach addresses each barrier.}
  \label{tab:barriers}
  \begin{tabular}{p{0.48\textwidth}p{0.48\textwidth}} % two column
  %\begin{tabular}{p{0.47\textwidth}p{0.47\textwidth}} % one column
    \toprule
    \textbf{Potential barriers to participation} &
    \textbf{How our approach addresses these barriers} \\
    \midrule
    \textbf{AI literacy}: Many design choices and assumptions that are baked into AI systems can be opaque to non-AI experts. & Our comicboards break down the AI development lifecycle to elicit specific feedback on different components of an AI system’s design. \\
    \midrule
    \textbf{Reading literacy}: Low reading literacy presents a significant barrier to co-design among many unhoused individuals. & Our comicboards combine illustrations with brief, carefully crafted captions to increase accessibility to individuals with lower literacy. \\
    \midrule
    \textbf{Social stigma}: There are significant stigmas surrounding homelessness. Unhoused individuals may be uncomfortable openly sharing their experiences and perspectives. & We create a gender-neutral persona that allows participants to self-determine when to bring in their own lived experiences and when to distance themselves from the discussion subjects at hand. \\
    \midrule
    \textbf{Power imbalance}: There are imbalanced power dynamics across our stakeholder groups (e.g. county workers versus unhoused individuals), which risk hindering safe and open conversation. & We conduct one-on-one sessions with participants, but share de-identified responses from prior study participants from other stakeholder groups, to facilitate asynchronous inter-group deliberation. \\
    \bottomrule
  \end{tabular}
\end{table*}

\subsection{Study Protocol}
We conducted AI lifecycle comicboarding in a series of one-on-one sessions with unhoused individuals and frontline workers in homeless services. We decided to run these sessions one-on-one, rather than in a workshop format, because we anticipated that participants in our study might not be comfortable sharing their experiences and viewpoints as openly in the presence of other participants \cite{holstein2019designing, stapleton2022imagining}. Each study session began with a brief interview portion to understand participants’ backgrounds. We then began the comicboarding activity to elicit participants’ feedback and ideas around the design of HAA, presenting one comicboard for each of eight major components of HAA’s design (as shown at the top of Figure \ref{lifecycle}). After participants provided their feedback, but before moving onto the next comicboard, we shared a few selected responses from other participants for discussion. This provided an opportunity for participants to express agreement or disagreement with others’ perspectives, or to build upon ideas generated by others, within the context of a one-on-one session. Given that these responses included design ideas from other stakeholder groups, this process also provided a safe space for deliberation between unhoused individuals, county workers, and external service providers, given the power dynamics among these groups (cf. \cite{holstein2019designing}). Finally, we revealed the final panel of the comicboard, describing how the relevant aspect of HAA is currently designed in reality. Participants were invited to provide feedback on the actual design choices that the developers of HAA had made, and to compare these with their own ideas, before moving onto the next comicboard. After going through all eight of the comicboards, we invited participants to reflect on their overall experience during the study, and we then wrapped up the study by collecting demographic information from participants.

\subsection{Recruitment}

\begin{table*}
  \caption{Participants’ self-reported demographics. Given the sensitive nature of our context, we present aggregated information.}
  \label{tab:demographics}
  \begin{tabular}{p{0.42\textwidth}p{0.54\textwidth}} % two column
  %\begin{tabular}{p{0.41\textwidth}p{0.53\textwidth}}
    \toprule
    \textbf{Demographic information} &
    \textbf{Participant counts or statistics} \\
    \midrule
    \textbf{Race} & Caucasian (12), African American (9)  \\
    \midrule
    \textbf{Age} & mean: 39.3, minimum: 26, maximum: 64 \\
    \midrule
    \textbf{Gender} & female (11), male (8), non-binary (2) \\
    \midrule
    \textbf{Homelessness Status} (unhoused individuals only) & currently unhoused (5), formerly unhoused (7) \\
    \midrule
    \textbf{Duration of Homelessness} (unhoused individuals only) & months (4), years (5), decades (3) \\
    \midrule
    \textbf{Years in the Field Unit} (county workers only) & mean: 2, minimum: 1.5, maximum: 3 \\
    \midrule
    \textbf{Services Provided} (service providers only) & street outreach (1), street medicine (1), education for youth (1), housing for women (1), housing for LGBTQ community (1) \\
    \bottomrule
  \end{tabular}
\end{table*}

We adopted a purposive sampling approach to recruit both frontline workers in homeless services and people with lived experiences of homelessness.

Through our contacts in county government, we recruited county workers in the field unit that mainly focuses on street outreach. Specifically, we first got into contact with a data analyst from the county's analytics team. Our contact then connected us with the field unit's supervisor, who shared our recruitment message with frontline workers in the field unit. These frontline workers have direct experience using HAA to run housing assessments on a daily basis, and they also have regular, face-to-face interactions with local unhoused populations.

Using contact information collected through public websites and databases of homeless services in the region, we also recruited non-profit service providers, whose services span street outreach, medical support, education for youth, and housing for women and the LGBTQ community. We connected with these service providers through emails, phone calls, or participating in local community meetings where community leaders connected us to trusted service providers they worked with. Because many of these service providers have built longstanding trust with local unhoused individuals, they are able to provide a unique birds-eye view of HAA's impacts on local unhoused populations in addition to insights into the system's impacts on their day-to-day work. They expanded on and extended the perspectives of the county workers and individuals with direct experiences of homelessness we interviewed.

Finally, we recruited participants with lived experiences of homelessness. These participants have on-the-ground knowledge of homelessness in the region and are directly impacted by HAA and/or prior assessment tools' decisions. Recognizing the ethical complexities of conducting research with unhoused individuals, we solicited feedback on recruitment strategies from county workers, service providers, and relevant domain experts. Based on their feedback, as well as prior recruitment practices in HCI research \cite{le2008designs}, we decided to recruit unhoused and previously unhoused individuals through county workers and service providers who have established relationships in the unhoused community as intermediaries.

In total, we recruited 21 participants, including 4 county workers, 5 non-profit service providers, and 12 people with current/former personal experience of homelessness. Recognizing that research sites can be a barrier to trust and acceptance among community members \cite{harrington2019deconstructing}, we provided several study locations, including both in-person and virtual options, for participants to choose from based on their preferences. In the end, we met all county workers in-person in the county's building; we talked to all service providers virtually over Zoom; we spoke with unhoused individuals either in-person on campus or virtually over Zoom or by phone. Each study session lasted 108 minutes on average, and all of our study participants were compensated \$60 for their participation. The amount of compensation was recommended by our contacts within the county who had extensive experience working with unhoused individuals and recruiting them for feedback sessions.

\subsection{Data Analysis}
To analyze our study data, we adopted a reflexive thematic analysis approach \cite{braun2012thematic, braun2019reflecting}. Three authors conducted open coding on transcriptions of approximately 38 hours of audio recording and generated a total of 1023 codes. Each transcript was coded by at least two people, including the researcher who conducted the interview for that transcript. Throughout this coding process, we continuously discussed disagreements and ambiguities in the codes, and iteratively refined our codes based on these discussions. Such discussions are critical in a reflexive thematic analysis approach, where different perspectives contribute to the collaborative shaping of codes and themes via conversation \cite{braun2019reflecting, mcdonald2019reliability}. Accordingly, in line with standard practice for a reflexive thematic analysis, we do not calculate inter-rater reliability, given that consensus and iterative discussion of disagreements is built into the process of generating codes and themes \cite{braun2019reflecting, mcdonald2019reliability}.

We also intentionally conducted our analysis \textit{across} comicboards, rather than conducting analyses per comicboard, given that our goal was to understand broader themes in participants' responses across the full set of comicboards. This analysis approach is common in HCI comicboarding and storyboarding studies \cite{moraveji2007comicboarding, davidoff2007rapidly}, and can be considered analogous to thematic analysis of results from interview studies, where coders do not necessarily organize codes based on specific prepared interview questions. Moreover, our participants sometimes returned to the comicboards they had already read and provided more feedback after learning more about how the system worked in later comicboards. This complicated the attribution of specific participant responses to specific comicboards. Many design choices in an AI system are inevitably intertwined, and participants responses across multiple comicboards reflected this interconnectedness.

After coding, we conceptualized higher-level themes from these codes through affinity diagramming. In total, this process yielded 65 first-level themes, 10 second-level themes, and three third-level themes. We present our results in Section \ref{result}, where section headers broadly correspond to second and third-level themes. All second and third-level themes are shown in Appendix \ref{themes}.

\subsection{Positionality Statement}
We acknowledge that our experiences shape our research, and our relative privilege within society provides us with advantages that our study participants do not hold. Specifically, we are researchers who work and receive research training in the United States in the fields of Human-Computer Interaction, Social Work, and Communication. Our team has prior research experiences in social work and public-sector technology. One author has direct work experience with unhoused communities and homelessness services. All authors live in the county where HAA is deployed. Two authors briefly experienced homelessness in the region but were never unsheltered on the streets. Another author had a family experience of homelessness when growing up outside of the region.

To conduct this research, we consulted domain experts in homelessness and worked closely with frontline workers in the county and non-profit organizations. Following prior literature \cite{oakley2013interviewing, reinharz1992feminist}, we also openly shared that we are researchers working independently from the county and agency that deploy HAA with participants before the interview and verified our interpretation of their quotes after the data analysis. Throughout this research, we followed prior approach in HCI \cite{starks2019designing, liang2021reflexivity, liang2021embracing} to pause and reflect on (1) Who would potentially benefit from the research outcome? (2) Are we truly supporting and serving the community? (3) What biases do we bring to this research? We paid particular attention to how the participants could directly benefit from participating in our study in addition to the compensation, considering that time is invaluable, especially in the case of unhoused individuals. We share reflections on our study from participants across all stakeholder groups in the Discussion (Section \ref{discussion}).

\section{Results} \label{result}
In this section, we organize our findings around three third-level themes identified through our analysis: (1) Desires for Feedback Opportunities, (2) Feedback on HAA’s Design, and (3) Feedback on HAA’s Deployment.

Overall, we found that \textbf{both unhoused individuals and frontline workers wished for opportunities to provide feedback on the design and use of the AI system}. As discussed in Section \ref{feedback}, participants noted the county had previously provided regular opportunities for feedback around older assessment tools. However, once the county adopted the ``black box’’ HAA, community members and frontline workers were no longer invited to provide such feedback. Within the county, there was significant skepticism that non-AI experts could provide meaningful feedback on an AI system. However, \textbf{our findings demonstrate that community members and frontline workers can provide specific, critical feedback on an AI system’s design and deployment, if empowered to do so}. 

As we will discuss in Sections \ref{design} and \ref{deployment}, participants shared concerns and design suggestions related to the AI system’s overall \textit{design objective}, specific \textit{model design} choices, the selection of \textit{data} used to train and run the model, and the broader \textit{sociotechnical system design} around the model’s deployment. Workers at the county’s field unit offered a unique perspective given their direct, day-to-day experience interacting with HAA, as well as their regular interactions with local unhoused populations. For example, county workers shared concerns and ideas for improvement based on their observations of systematic limitations and errors in HAA’s predictions. Complementing county workers’ perspectives, external service providers brought a birds-eye view of the system’s \textit{downstream} impacts on both unhoused individuals and service providers in the region. Finally, unhoused individuals brought in direct lived experience of homelessness, and were able to compare the HAA’s design and deployment against their own on-the-ground knowledge of homelessness in the region.

Throughout this section, county workers are identified with a ``W,’’ non-profit service providers are identified with a ``N,’’ and unhoused participants are identified with a ``P.’’

\subsection{Desires for Feedback Opportunities} \label{feedback}
Both service providers and county workers shared that there used to be a more open community feedback process around previous assessment tools. However, after the county adopted the ``black box’’ HAA, they perceived that workers and community members were no longer invited into conversations to provide such feedback. All participants expressed a desire for regular opportunities to give feedback on the system’s design and deployment, for example through regularly-held community feedback sessions. Although the county already held regular feedback sessions related to their programs and services, these sessions typically avoided topics that were assumed to be overly technical. 

Participants shared that, in the past, they were able to discuss and update the questions used in the county’s previous, questionnaire-based assessment tool. For example, a service provider noted that \textit{``one of the things with [the questionnaire-based tool], you know, at least once a year, we were able to get into conversations about whether there were questions that we felt would be relevant to better understand some of these risks to be added to the system. We don't get that opportunity now''} (N2). Although some light consultations, such as focus groups and information sessions, occurred before HAA's deployment, county workers and service providers perceive that there has been no follow-up. For example, a county worker said \textit{``I think it would be helpful to bring [unhoused individuals] back in the room. [...] I do remember when we were bringing people in to discuss it, but I didn't hear anything about follow up after that''} (W3). Similarly, a service provider recalled that \textit{``prior to [HAA]’s deployment, there is a forum that is facilitated by [the county]. [...] There were a couple of sessions introducing the algorithm. But after that point, we never had any kind of follow up’’} (N3). 

All of our participants expressed a desire for continuous, post-deployment feedback channels around HAA. However, they perceived that frontline homeless service workers and people with lived experiences of homelessness are insufficiently involved in the design and evaluation process for tools like HAA: \textit{``Most times people that do the kind of outreach work, where they're face-to-face with community members and doing the hardest lift, they're the least sought out as far as like research is concerned, and that should be the exact opposite’’} (N1). Participants also emphasized that it is critical to get direct feedback from people with lived experiences of homelessness, who are directly impacted by HAA’s decisions: \textit{``I think [the unhoused individuals] should be allowed to be involved because this is about them. They should be allowed to be heard. Not just the staff, not the county. They get to go home and sleep at night’’} (P10).

Some service providers worried that the county currently intentionally avoids gathering legitimate feedback about HAA or discussing potential flaws with the system: \textit{``I think the county doesn't actually get legitimate feedback on housing programs. It reports feedback based on predetermined criteria, but they intentionally don't do that effectively’’} (N2). Meanwhile, county workers shared that there was significant skepticism within the county that community members or other non-AI experts could provide meaningful feedback on a complex system like HAA: \textit{``I don't think you're gonna get feedback from [them] about HAA, or the assessment, or anything like that. I think [people can give feedback based on their] experiences of the housing program’’} (W2).

\subsection{Feedback on HAA’s Design}\label{design}

As participants gained insight, through our comicboarding method, into HAA’s design and use, they raised a number of concerns and suggestions for alternative designs. Reflecting on their experience during the study, several participants noted that \textbf{our comicboarding approach helped to open up conversations around otherwise hidden assumptions and decisions underlying the algorithm’s design and deployment}. As we discuss below, participants questioned the algorithm’s overall design objectives, aspects of its model design, and the selection of data used to train and run the algorithm. For example, as participants learned more about HAA’s overall design, they expressed concerns about ways the algorithm may be optimizing more towards the county’s interests rather than the actual needs of unhoused individuals (Section \ref{objective}). Participants also expressed concerns about the validity and reliability of specific aspects of the model’s design, such as the current choices of proxy variables used to measure particular real-world outcomes (Section \ref{model}). Finally, participants shared real-life examples to illustrate how the algorithm’s use of incomplete, decontextualized, or potentially misleading data may lead to erroneous scores and disadvantage particular populations (Section \ref{data}).

\subsubsection{Feedback on HAA’s objective} \label{objective}
\textbf{Some participants challenged HAA's problem formulation – the very idea of prioritization}. They believed that \textit{``everyone has the right to housing just like a survival need’’} (P2). A currently unhoused participant raised the question to the county: \textit{``what more do you need for me to tell you that I'm important enough to live somewhere’’} (P12). Participants suggested the county focus its resources on addressing the root problems rather than prioritization, such as providing more housing or preventing homelessness in the first place: \textit{``[The county] could get [homelessness] end more effectively, if they were able to take in people that had just hit homelessness, instead of they have to go through god knows what and see if they even survive it’’} (P3).

In addition, as participants learned more through our comicboards about HAA’s overall objectives and the specific outcomes it predicts, they \textbf{perceived that HAA had been designed and evaluated to reflect the county's values and serve their interests rather than to reflect local community members' needs}. Even when participants acknowledged that it may be necessary to prioritize housing resources, given that the system is currently set up in a way that guarantees resource scarcity, they raised concerns with the particular ways HAA implements prioritization. For example, a formerly unhoused participant said: \textit{``It looks like they're trying to [prioritize] people that are causing problems for others in society, as opposed to people that are at risk for themselves [and] are suffering internally’’} (P2). A service provider shared a similar concern: \textit{``Mental health inpatient and all of these things have significant financial tags associated with them when a person experiencing homelessness. [...] It's possible that these are the best proxies that exist, but it looks like we're measuring more financial cost to systems that have power, than we are measuring actual harms to actual people’’} (N5).

Meanwhile, several participants expressed \textbf{concerns that the metrics used to evaluate whether HAA is successful seem to encode specific values that are not focused on the subjective experiences and outcomes of unhoused individuals}. Instead, a service provider suggested evaluating HAA based on more client-focused outcomes: \textit{``The best metric for success is going to be: did the people who went through the system and received services have better outcomes than they were having before the system was implemented? [...]  Do they maintain stable housing? Do they find employment that is sustainable and satisfying?’’} (N5). They questioned who had been involved in or excluded from determining the evaluation metric: \textit{``What does it mean when something performs better. [...] Whose version of better is that?''} (W3).

\subsubsection{Feedback on HAA’s model design} \label{model}
Participants voiced several concerns about the validity of the HAA model. First, \textbf{some participants were skeptical about whether it is truly possible to predict a person's likelihood of being harmed on the streets, based on their administrative records}. For example, a formerly unhoused participant said: \textit{``There's just no way for a computer to accurately predict somebody's future based on that limited data. [...] If a person is alone, scared, and suffering from fears of institutions, a computer doesn't know that’’} (P2). Similarly, a service provider worried that people can be more vulnerable on the streets for many reasons that may not be reflected in administrative records: \textit{``Some people are much more likely to be victimized on the street, [...], that may or may not really get picked up in the data. People who have patterns of behavior, or relationships that are recurrent, that get them into a relationship with an abuser, or they're taken advantage of physically, sexually, economically’’} (N4).

\textbf{Participants were particularly concerned that the proxy outcomes that the HAA model is trained to predict (e.g., counts of hospital visits and stays in jail), while conveniently available in administrative data, could be highly misleading}. For example, some county workers expected that some people may visit the emergency room in order to get out of the cold, not because they are truly sick. Similarly, as W2 noted: \textit{``If a person is, over the past two years, in the emergency room every week, but they drop off, and then we find that they have moved out into a tent. [...] Their health is probably worse than it was when they were going to the emergency room every week''} (W2). Without data that could provide insight into the actual \textit{causes} behind these observed outcomes, participants worried that HAA’s predictions might lead decision-makers astray: \textit{``I don't know how a computer considers whether a person will continue to experiencing homelessness if we don't understand what brought him there in the first place. [...]  I don't think we have the right data. We're looking at outcomes and not the causal effects of what resulted in that outcome’’} (N3). 

Given these concerns, even though participants generally agreed that the housing allocation process should \textit{``ideally see people more vulnerable floating to the top of the list and getting served more quickly’’} (W1), \textbf{they were concerned that the score generated by HAA’s model could not accurately reflect who was in more urgent need of help}. For example, a service provider argued that \textit{``somebody with [...] multiple years of chronic homelessness is far safer on the street than a 25-year-old white female newly on the street alone, [... but] this individual would return like a zero or one’’} (N2). They perceived that the current prioritization process \textit{``basically says, until you can experience the real depth of the traumas associated with living on the street, you're gonna have to stay out there’’} (N2). Participants suggested that there should be multiple pathways for prioritization because \textit{``we'll find folks that are pretty stable, but they just need an extra pick-me-up. It doesn't always feel like there are a lot of resources for those people. So I think there should be two paths’’} (W3).

\textbf{Participants expressed conflicting views regarding whether the HAA model should account for identity characteristics, such as ethnicity.} For example, after learning from our comicboards that HAA does not consider race when making its predictions, some participants brought attention to the vulnerabilities that these characteristics introduce: \textit{``statistically speaking, Black people are discriminated against every day of their life on a macro and micro level, both professionally and personally. [...] How can you claim something is equitable, when you're not even considering race?’’} (N1). Meanwhile, several unhoused participants argued for a demographic-blind prioritization process. For example, a currently unhoused participant who self-identified as African-American argued that \textit{``I really don't think race is a big deal because everybody has their battles. [...] No matter what race we are, anything can happen to us’’} (P4).

Finally, participants worried that HAA's predictions may become less and less valid as time progresses because, although the HAA model is static, they \textbf{believed the true relationship between the model’s inputs and the real-world outcomes that it predicts is highly unstable across time} (cf. \cite{quinonero2008dataset}). A county worker, W2, offered an example to illustrate how a change in policy or the launch of a new program could impact the validity of the model’s predictions. W2 noted that after a new program, the \textit{Continuum of Care} was initiated, their organization expanded their efforts to proactively find unhoused individuals: \textit{``So I think the past profiles are not a perfect correlation to the present and the future because [...] we were housing people who were calling us on the phone. The vulnerability changes as we become better at reaching people who are more vulnerable’’} (W2). Another participant echoed this concern that HAA is trying to model a rapidly shifting target, yet it is trained on outdated data: \textit{``The living cost, income, nothing is the same. How can you predict for now when nothing in the world is the same? [...] The world evolves every day, so their criteria for the prediction should evolve with the world’’} (P7).

\subsubsection{Feedback on HAA’s data} \label{data}

As described in Section \ref{studycontext}, HAA relies on administrative data in the county’s data warehouse. \textbf{Participants worried that HAA misses the data of people who are averse to accessing public services due to institutional violence and prior traumas with these systems.} As a county worker argued: \textit{``the lack of engaging in a service is not a lack of need’’} (W2). A formerly unhoused participant shared his experience of institutional violence: \textit{``I was in a lot of pain and I couldn't really move. [...] I went to the hospital. [...] They searched me when I went in. [...] They came and searched for me again. [...] They brought me upstairs and searched me a third time. So I left. [...] They were looking at my record and judging me based on my past experience at the hospital that I was likely to have drugs on me’’} (P2). Another participant shared a similar experience and how it delayed her cancer treatment: \textit{``I had cancer inside my body. That was misdiagnosed because they thought I was there for other reasons to get free opioids’’} (P7). In light of such fears and mistrust around public services, a county worker suggested that sometimes \textit{``the drop-off is a greater indicator of the vulnerability than the continued engagement with the service’’} (W2).

\textbf{Participants also worried that the use of HAA could further disadvantage certain populations}, as they anticipated additional causes of systematic missingness in the data. For example, a county worker (W3) and a service provider (N3) mentioned that medical outreach and free clinics do not keep track of people's medical records. They also argued that young adults and people new to homelessness may not have records on file. They wondered whether HAA’s current design was basically saying: \textit{``because you haven't been homeless long enough, you don't deserve housing’’} (N3). Furthermore, participants mentioned that people who suffer from domestic violence might provide false information about themselves: \textit{``because of the fear of their perpetrator finding them’’} (N3). Finally, participants expressed concerns that: \textit{``some folks who are really struggling with their mental health might not realize how vulnerable they are’’} (W3). For example, a medical service provider shared that \textit{``I'm much more concerned with people that are eating out of garbage cans, that are not admitted to mental health hospitals; people who are sleeping next to busy intersections so that the noise will drown out the voices in their head, who are not going to inpatient admissions’’} (N4).

Furthermore, \textbf{participants were concerned that the administrative records upon which HAA relies would often be outdated, failing to reflect significant changes in people's situations}. For example, a participant shared that she was likely to face homelessness again, but for a completely different reason than her prior experience of homelessness: \textit{``I was facing totally different issues than I am now. The reason for me needing them before is because I had no support. I had no family. I was a foster kid. Now, I'm an adult. I can't work. I'm facing health issues. So my issues back then weren't what they are today’’} (P7). P7 shared with us that she tried to call the county to update her information but couldn't reach anyone on the other end. Participants also noted that a person's situation could change rapidly, even within a few days: \textit{``that was my biggest concern because somebody's situation could change over a weekend. [...] The incidence of violence or financial background, that kind of stuff can change pretty quickly’’} (N1).

Finally, participants emphasized that, \textbf{to complement available quantitative data, it is critical to consider qualitative narratives when making decisions about housing prioritization}. They argued that simply increasing the granularity of the numeric features currently used by the model (e.g., trying to capture broad categories of reasons why a person went to jail) would never be able to replace the need to consider such narratives. For example, a service provider suggested: \textit{``having a qualitative narrative of what those things were, I think that might be better than trying to assign a numerical value to a pot charge versus a domestic violence charge’’} (N5). In addition, participants believed that other important factors, such as social dynamics on the street, could only be captured through qualitative narratives: \textit{``you're not really accessing some of the social dynamics that aren't digitized. They're more of a narrative of story. The situation on the streets is like following a soap opera [...] because people are relating to each other. There are dangerous domestic violence situations, people that are threatening to kill other people, and people who are beginning to give up and overdosing themselves on purpose. I don't think that a lot of those types of data, either medical, psychiatric, or soap opera are going to be captured by the computer’’} (N4). In order to capture such narratives, participants argued for the importance of human investigation, to complement analyses of quantitative data.

\subsection{Feedback on HAA’s Deployment} \label{deployment}
Frontline workers both within and outside of the county shared experiences where they had observed potentially harmful errors in HAA’s scoring behavior. When they encountered such behaviors, workers often felt powerless to take action—particularly given that information about how HAA works is intentionally withheld from them. Although county workers shared ways they are currently able to exercise their discretion and advocate for individuals to be prioritized differently, they noted that they were discouraged from using these mechanisms. Participants felt strongly that frontline workers should be empowered and extensively trained to override HAA's recommendations when appropriate (Section \ref{currentdesign}). They also suggested broader changes to the sociotechnical system around HAA's deployment (Section \ref{broaderdesign}), and proposed alternative ways of using data to streamline provision of services (Section \ref{streamlinedesign}).

\subsubsection{Feedback on the current deployment} \label{currentdesign}

Throughout our study, \textbf{frontline workers both within and outside the county shared multiple examples of what they believed to be erroneous, potentially harmful algorithmic behaviors. However, they felt frustrated and powerless when they encountered these situations}: \textit{``When someone’s score doesn’t qualify them for anything, I kind of have to share that [with unhoused individuals]. And then, it’s just we’re both helpless in that situation’’} (W1). County workers noted that even though they directly interact with HAA day-to-day, they were kept from knowing how the algorithm calculates the scores or weighs different features: \textit{``[My supervisor] always says, [the model developers] tell him don't ask what's in the sausage.’’} (W1). In the absence of formal insight into how scores are computed, workers developed their own hypotheses about how the system works and where the system may be less reliable, based on their daily interactions with the system (cf. \cite{kawakami2022improving}). 

Workers shared that, in cases where the score does not match a person's representation, they sometimes try to override HAA with alt HAA or submit prioritization requests. However, they are discouraged from using these override mechanisms. For example, a county worker shared that when they disagreed with HAA’s score, \textit{``the only other thing you can do is run the alt HAA, and we're really not supposed to do that’’} (W1). Meanwhile, some unhoused participants found the questionnaire-based alt HAA itself easily gamifiable: \textit{``I'm a really good test taker. That's how I got in’’} (P3). Aside from alt AHA, prioritization requests heavily rely on individual advocacy, which overloads individual workers without systematic support. For example, a county worker shared an experience submitting a prioritization request on behalf of a person who was diagnosed with schizophrenia and stayed in abandoned houses for two years but only scored a three: \textit{``I have to work really hard to be able to offer this person services [...] because he's not going to be served through this system’’} (W1). 

Considering these limitations, \textbf{participants suggested that a system like HAA should at least blend the unique strengths of human judgment with those of data-driven algorithms}. For example, a participant suggested the county \textit{``should continue [the system] but take some of the information that [comes out of] this research and update the system and have more human involvement in the calculations’’} (P7). Participants also believed that frontline workers who use HAA should be empowered and extensively trained to know when to rely on HAA versus when to override HAA's recommendations: \textit{``The person should ultimately be responsible for the decision. They should go through extensive training on how to make those decisions and how to weigh the scores versus their intuitions and the new information that the client has given’’} (N5). Despite these suggestions, unhoused participants still expressed worries that the county may over-rely on HAA and use it as an excuse to avoid more time-consuming but valuable investigations.

\subsubsection{Feedback on the broader sociotechnical design considerations} \label{broaderdesign}

In addition to human intervention, participants also asked for broader changes to the sociotechnical design around the algorithm's deployment. Specifically, they argued \textbf{more upstream work is needed to connect people to HAA in the first place, and downstream care plans are required for people to succeed after receiving a score from HAA.} 

Participants argued there is upstream work needed to connect people to HAA because many unhoused individuals who don't utilize services are flying under the radar. As a medical service provider put it: \textit{``You shouldn't just be worried about who's in the waiting room. You should be worried about who's not in the waiting room’’} (N4). Participants shared various reasons why the unhoused individuals do not connect with HAA in the first place. For example, people may not have the inclination to reach out because of their distrust in institutions: \textit{``People don't want to go [to the county] because of the trust issue’’} (P3). In addition, some people are too vulnerable to seek resources on their own: \textit{``It’s weird that you got to prove that you're homeless [to get housing]. Most people aren't in the mental state to be able to prove anything in my experience’’} (P3).

Participants also suggested that downstream care plans should be customized for each individual to prevent them from cycling back to homelessness. For example, a formerly unhoused participant shared that \textit{``I actually have a friend. He's in a house, [but] he can't stop going out and panhandling and flying signs. [...] Moving into a house doesn't make you not in a homeless mind state’’} (P3). County workers also shared examples where people who receive high scores are assigned to programs with a lower level of support and have traumatic experiences in those programs: \textit{``The big problem we saw last year is that some 10s are getting rapid programs because there is more availability of resources for the lower level of support. [...] There are people experiencing some trauma from getting into those housing programs’’} (W2). Participants suggested there should be a systematic effort to follow through and help a person succeed in the end so that \textit{``all this information about the vulnerability isn’t just their ticket into the program door’’} (W2).

\subsubsection{Ideas for alternative uses of data, to streamline services}\label{streamlinedesign}

\textbf{Participants shared multiple ideas for alternative ways to use data to streamline service provision, which they perceived as more valuable than the current design of HAA.} For example, several unhoused participants suggested that the system can actively connect people to various resources, not only housing, based on the data already available to the county: \textit{``I think that's the way a computer can help because they'll weave out what is needed for you to best suit your needs’’} (P8). County workers also asked for better information exchange about unhoused individuals to form individualized care plans and alleviate their burden: \textit{``There's so many times where the person who's being served has to sit down and have the same conversation and reveal the same information over and over again. Why isn't the computer doing a better job of alleviating that burden from the person’’} (W2). Finally, service providers noted that changes to HAA’s primary data sources within the county's data warehouse, such as the Homeless Management Information System (HMIS), could have an enormous impact: \textit{``HMIS is something that is part of the federal government requirements for everybody that receives funding from HUD, [...] but it's archaic, horrible, poorly designed, and rarely updated’’} (N2).

\section{Discussion} \label{discussion}
Given the spread of ADS systems in homeless services, it is critical to understand directly impacted stakeholders' perspectives on these systems. In this paper, we present the first in-depth understanding of frontline workers' and unhoused individuals' desires and concerns around the use of AI in homeless services. To elicit feedback on a deployed ADS system from stakeholders spanning a wide range of relevant literacies, we employed \textit{AI lifecycle comicboarding}: a feedback elicitation and co-design method that adapts the comicboarding method \cite{hiniker2017co, moraveji2007comicboarding} to scaffold both broad and specific conversations around different components of an AI system’s design, from problem formulation to data selection to model definition and deployment. In this section, we highlight key takeaways, reflect on our experience using AI lifecycle comicboarding, and discuss considerations for the use of this method in future research.

As discussed in Section \ref{feedback}, within county government, there was significant skepticism that stakeholders without AI expertise could provide meaningful feedback on the design of an AI system. Given this skepticism, community members and frontline workers were given minimal opportunities to learn about the HAA system or provide feedback. Yet our findings suggest that non-AI experts can provide specific, critical feedback on an AI system’s design and use, if invited and appropriately empowered to do so. As our participants gained insight, through our comicboarding method, into HAA’s design (Section \ref{design}) and deployment (Section \ref{deployment}), they raised a number of concerns and suggestions. Some of participants’ design feedback surfaced broad concerns. For example, as participants learned more about HAA’s overall design, they expressed concerns about ways the algorithm may serve to optimize more towards the county’s interests rather than the actual needs and safety of unhoused individuals (Section \ref{objective}). Participants also shared ideas for alternative ways to use the data currently available to the county, which they perceived as more likely to bring benefits and less likely to cause harm, compared with HAA’s current design (Section \ref{streamlinedesign}). In addition, participants provided specific feedback on particular \textit{model-level} and \textit{data-level} design choices. For instance, participants expressed concerns about the validity and reliability of specific aspects of the model’s design, such as the current choices of proxy variables used to measure particular real-world outcomes (Section \ref{model}). In light of the limitations of available administrative data (Section \ref{data}), participants also offered additional deployment suggestions, including ways to elevate human judgment in the decision-making process (Section \ref{currentdesign}) and ways to improve the design of the broader sociotechnical system surrounding HAA’s deployment (Section \ref{broaderdesign}).

Reflecting on their experiences during the study, participants noted that our comicboarding approach helped to open up conversations around hidden assumptions and decisions underlying the AI system’s design and deployment, which would have otherwise remained opaque to them. For example, a service provider (N4) noted that although they were aware of HAA prior to the study: \textit{``I think it helped me understand [how HAA works] because I only knew [in] general terms, I didn't really see the process laid out like that.’’} N4 shared that, after going through the full set of comicboards: \textit{``It also made me concerned about the limitations of you know, garbage in garbage out. Or incomplete, incomplete out.’’} Similarly, N3 reflected that, \textit{``[The] user friendly narrative around the storyboards was effective for me. It helped me better organize my thoughts. [...] I come out of that [...] with a much better understanding of the HAA.’’} N3 added that following their experience in the study, \textit{``The validity of [HAA] causes me great concern.’’} These concerns motivated some service providers to reach out to the county after the study with the desire to improve the system and mitigate the potential harm they identified through our comicboards.

Meanwhile, participants also reflected on how their participation in the study had direct benefits beyond monetary compensation. For example, after understanding how HAA currently generates scores based on personal information, some unhoused individuals decided to actively reach out to the county in order to update their life situations: \textit{``It gave me a better understanding for the application process, I didn't know how they went about that at all. It let me know certain things that I can do on my end to help my chances of getting the housing. After this, I'm gonna try to reach out and provide updated documentation for my medical issues}’’ (P7). For one county worker, who interacts with HAA day-to-day, participating in the study made them feel empowered to explain how HAA’s algorithmic decisions are made: \textit{``This is revealing to me that I can be like, more upfront with people when I do give them the score and let them know, this is exactly what you are, you know, what you're eligible for [...] I think it would be good [...] for me to be a little bit more transparent with people’’} (W1).

In this study, we used AI lifecycle comicboarding to solicit feedback and suggestions for design modifications to an AI system \textit{after} that system had already been deployed. While collecting post-deployment feedback is critical, as discussed in Section \ref{feedback}, we also encourage using it early in the design process of a new AI system (e.g., conceptualization, prototyping, or pilot testing stages). Indeed, at later stages, once significant resources have been invested in an AI system’s development, there is a risk that system developers will be hesitant to implement broader changes such as those discussed in Section \ref{objective} and Section \ref{streamlinedesign}, and may instead be biased towards more incremental changes that can be implemented with fewer resources. When using the method before a system is already in place, the final panel of each comicboard (e.g., Figure \ref{fig:teaser} (e)) can be used to elicit feedback on \textit{proposed} designs. With this flexibility, our method is intended to empower participants not only to critique and redesign existing AI systems, but also to redirect the design processes of proposed systems at the earliest stages of the design process.

In the future, more formal and controlled evaluations of AI lifecycle comicboarding would help further understand its strengths and limitations in elicting specific feedback compared to standard interviews or alternative comicboarding approaches. We envision that our method could be adapted and used to elicit feedback from other stakeholder groups who experience intersectional social disadvantages, such as children and families subject to out-of-home placements and state intervention or individuals with intellectual/developmental or psychiatric disabilities. In future work, we also plan to develop a suite of toolkits based on the comicboards generated for this study, along with instructions on how to adapt and use them. We hope that releasing these toolkits will help HCI practitioners and developers of AI systems to better integrate the voices and perspectives of impacted stakeholders.

\section{Conclusion}
Our study demonstrates that with an appropriate feedback elicitation method, community stakeholders spanning diverse backgrounds and literacies can provide specific and critical feedback on an AI system’s design. Using our method, we have presented an in-depth understanding of frontline workers’ and unhoused individuals’ perspectives on the use of AI in homeless services. Future research should explore ways to adapt the AI lifecycle comicboarding method for use with other stakeholder groups, who may face additional barriers to participation in AI design and critique. In addition, future work should explore the design of practical processes, policies, and technical approaches that can support the effective \textit{incorporation} of stakeholder feedback into AI system design in practice.

% Acknowledgements
\begin{acks}
We thank our participants for their time and input that shaped this research. We also thank our contacts in the county and non-profit service providers for helping with the recruitment and verifying the comicboards. Finally, we thank Laura Dabbish, Yodit Betru, Bonnie Fan, Jordan Taylor, Wesley Deng, Logan Stapleton, Anna Kawakami, Jane Hsieh, Seyun Kim, Tiffany Chih, and anonymous reviewers for their insightful feedback on the study design and paper draft. This work was supported by the National Science Foundation (NSF) under Award No. 1939606, 2001851, 2000782 and 1952085, and the Carnegie Mellon University Block Center for Technology and Society (Award No. 55410.1.5007719).
\end{acks}

\bibliographystyle{ACM-Reference-Format}
\bibliography{submission}

\appendix

\section{Comicboards}\label{comicboards}
\begin{figure*}[!ht]
  \centering
  \includegraphics[width=0.69\linewidth]{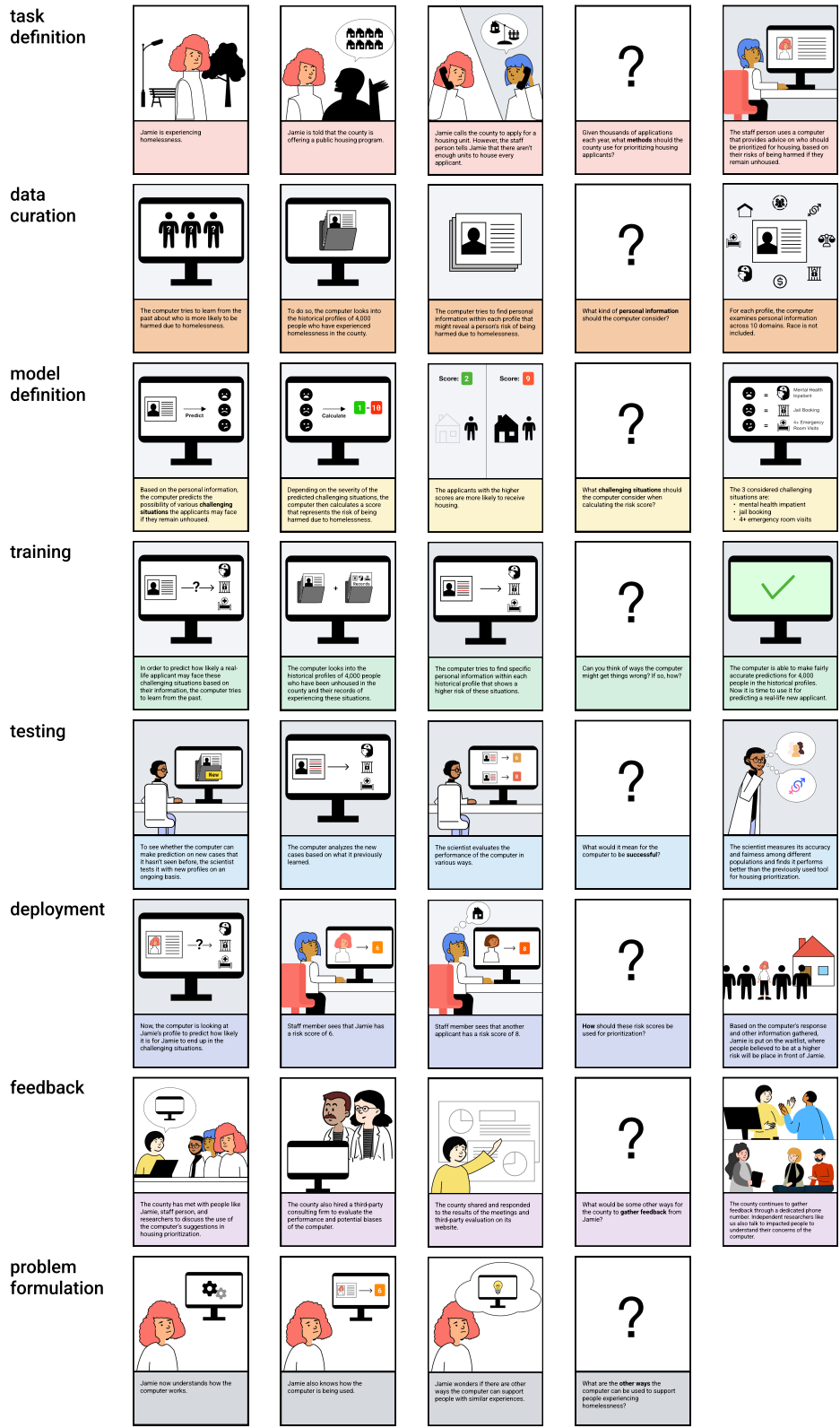}
  \caption{The comicboards we developed and used for our study.}
  \Description{This figure has all the comicboards we developed. There are eight rows in this figure. Each row corresponds to the eight major components of an AI system's design and has a comicboard. Each comicboard has five panels. For example, the first row corresponds to the task definition and has the identical comicboard to Figure 1.}
  \label{fig:comicboard}
\end{figure*}

We include all the comicboards we developed and used for the study in Figure \ref{fig:comicboard}. Each row corresponds to one of eight major components of an AI system's design. While problem formulation is typically the first stage within AI's development lifecycle, we showed the corresponding comicboard at the end of our study in order to broaden the discussion once participants had a better understanding of the current system. We also kept the last panel of this comicboard open-ended to invite a broad range of alternative problem formulations. This presentation order is a design choice, given that HAA has been deployed for over two years. When using our method in contexts where an AI system is at earlier stages of conceptualization, researchers are encouraged to explore alternative presentation orders (e.g., starting with the problem formulation) as appropriate.

\section{Higher-Level Themes}\label{themes}
We provide a summary of the higher-level themes we identified through a reflexive thematic analysis approach in Table \ref{tab:themes}. These themes broadly correspond to the section headers in Section \ref{result}. Due to the limitation of word counts and space, we do not include the 65 first-level theme and 1023 codes in the table.

\begin{table*}
  \caption{The three third-level themes and ten second-level themes we identified through data analysis.}
  \label{tab:themes}
  {
  %\begin{tabular}{p{0.48\textwidth}p{0.48\textwidth}} % two column
  \begin{tabular}{p{0.3\textwidth}p{0.66\textwidth}} % one column
    \toprule
    \textbf{Third-level themes} &
    \textbf{Second-level themes} \\
    \midrule
    desires for feedback opportunities 
      & perceptions of a more open feedback process for previous assessment tools \\
      & perceptions of insufficient community involvement in HAA \\
      & desires for continuous, post-deployment feedback channels \\
      & perceptions of the county's intention of gathering legitimate feedback\\
    \midrule
    feedback on HAA's design
      & feedback on HAA's objective \\
      & feedback on HAA's model design \\
      & feedback on HAA's data \\
    \midrule
    feedback on HAA's deployment
      & feedback on the current deployment \\
      & feedback on the broader sociotechnical design considerations \\
      & ideas for alternative uses of data to streamline services \\
    \bottomrule
  \end{tabular}
  }
\end{table*}

\end{document}